**The three-dimensional impulse-response model: Modeling the training process in accordance with energy system-specific adaptation**

Hilkka Kontro[1], Armando Mastracci[2], Stephen S. Cheung[3], and Martin J. MacInnis[1]

1 Faculty of Kinesiology, University of Calgary, Canada

2 Baron Biosystems Ltd., Toronto, Canada

3 Department of Kinesiology, Brock University, Canada

ORCID: 0000-0001-6999-3096 (HK); 0000-0002-6149-4978 (SSC); 0000-0002-0377-5800 (MJM)

**Corresponding author:**

Hilkka Kontro

University of Calgary

2500 University Drive NW

Calgary, AB T2N 1N4

Canada

Email: hilkka.kontro@ucalgary.ca

Phone: +1 (403) 220-3841

**Running title:** 3D impulse-response model of training

**Word count:** 6008

ABSTRACT

Athletic training is characterized by physiological systems responding to repeated exercise-induced stress, resulting in gradual alterations in the functional properties of these systems. The adaptive response leading to improved performance follows a remarkably predictable pattern that may be described by a systems model provided that training load can be accurately quantified and that the constants defining the training-performance relationship are known. While various impulse-response models have been proposed, they are inherently limited in reducing training stress (the impulse) into a single metric, assuming that the adaptive responses are independent of the type of training performed. This is despite ample evidence of markedly diverse acute and chronic responses to exercise of different intensities and durations. Herein, we propose an alternative, three-dimensional impulse-response model that uses three training load metrics as inputs and three performance metrics as outputs. These metrics, represented by a three-parameter critical power model, reflect the stress imposed on each of the three energy systems: the alactic (phosphocreatine/immediate) system; the lactic (glycolytic) system; and the aerobic (oxidative) system. The purpose of this article is to outline the scientific rationale and the practical implementation of the three-dimensional impulse-response model.

## I. INTRODUCTION

Of all means to improve athletic performance in humans, the most effective method is physical training. While this is a self-evident and universally accepted fact, substantial uncertainty remains around the type, volume, and intensity of training that athletes should complete to optimize performance for a given event (Burnley et al., 2022; Foster et al., 2022; Laursen, 2010) and the periodization approach athletes should follow to peak for competition (Issurin, 2010; Kiely, 2010). Nevertheless, the main purpose of physical training is to induce physiological changes that result in more speed, strength, or fatigue-resistance, and depending on the specific goal of training, different modes and volumes of training are prescribed.

### *Principles of adaptation*

The improvements in performance following training are explained by the adaptive stimulus induced by exercise, whereby physiological systems respond to acute challenges in maintaining homeostasis (Holloszy & Coyle, 1984; Scheuer & Tipton, 1977). When the stress experienced is high enough, it elicits an adaptive response in the form of altered gene expression and subsequent protein synthesis. The magnitude of this stress is the impulse that determines the magnitude of the adaptive response. The effect of these adaptive responses accumulates over time to produce functionally and structurally enhanced biological processes that, for example, facilitate a higher maximal or sustained energy turnover during muscular contractions (Egan et al., 2013). Disruptions in homeostasis are greater the closer to a performance ceiling an athlete is exercising (Black et al., 2017), meaning that higher intensities and longer durations generally lead to greater adaptive responses (Fiorenza et al., 2018; Inglis et al., 2024; Montalvo et al., 2022). Depending on the task, a performance ceiling may exist due to a rate limitation or a capacity limitation in energy provision (Morton, 2006). The importance and consequences of this distinction in training modeling will be discussed in later sections.

### *Principles of modeling performance*

Exercise physiology often employs a reductionist approach to explain how performance can be pinpointed to the primary components of the human body. However, a holistic approach may also be adopted, in which the body is seen as an emergent system and performance arises from the integration of its primary components (Larina et al., 2022). In line with this approach, mathematical models have been proposed to explain the changes in performance following training and cessation of training. In a systems model, performance (the output or response) can be predicted from the training completed (the input or impulse). Banister and Calvert were the first to introduce and validate a systems model to explain how performance changes in response to adjustments in training load (Banister et al., 1975; Calvert et al., 1976). After some initial models involving multiple components, they reduced their performance model to be a function of two variables: fitness and fatigue (Banister & Calvert, 1980; Calvert et al., 1976). This impulse-response model, often referred to simply as the Banister model, survives to the present day and is, in various forms, the basis for current approaches to monitoring and predicting fitness, fatigue, and performance.

The theory behind the Banister impulse-response model builds on the observation that while increases in training lead to improved fitness (the positive component), they also generate fatigue (the negative component). Performance at a given time point thus equals fitness minus fatigue (Figure 1).

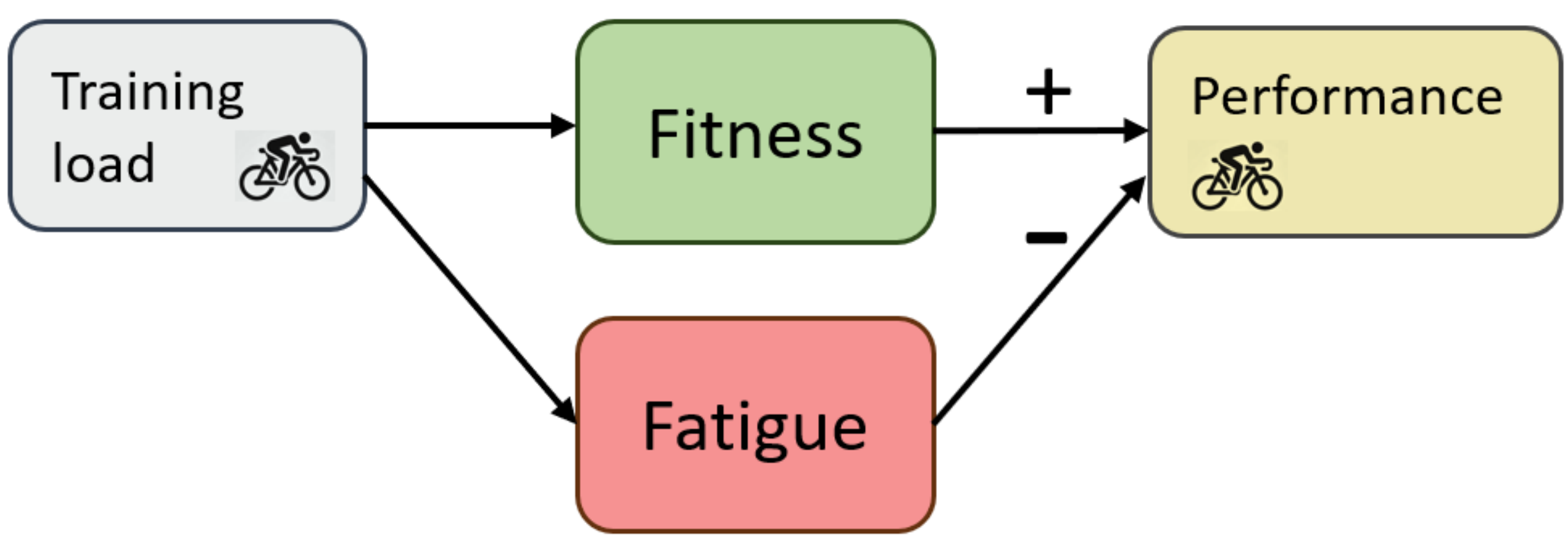

Figure 1. The Banister impulse-response model as redrawn from Morton et al. 1990. Performance is a function of fitness and fatigue, which both increase in response to training, but by different magnitudes and following different timeframes.

## II. A NOVEL POWER-BASED TRAINING LOAD METRIC

In order to apply an impulse-response model to training, one of the fundamental premises is a valid metric for the input. Training load describes the dose of training and is the required input to model the dose-response relationship of training and adaptation. In this section, we introduce a novel way of analyzing power output-based training data.

### *How can training load be quantified?*

Training loads can be classified into external/absolute loads or internal/relative loads. External measures of load include objective measures such as duration, work, work rate, speed and distance. Conversely, internal loads reflect the training-induced stress experienced by the athlete (Halson, 2014). While external training load is easily quantifiable, particularly in cycling with the widespread use of power meters, it is the magnitude of the internal load that is thought to initiate the adaptive responses to training (Impellizzeri et al., 2019). Internal loads can be assessed using physiological and perceptual markers or by anchoring an external measure of work rate to individual exercise thresholds. In the following section, the existing methods will be briefly discussed.

### *Popular models for heart rate-derived training load*

To quantify the dose of training based on heart rate (HR), a training load metric called training impulse (TRIMP) was introduced by Banister and colleagues (Morton et al., 1990). It factors in both duration and HR relative to the individual's HR reserve (maximal HR – resting HR). Briefly, Banister's TRIMP is calculated from duration of exercise and the (expected) relationship between HR and blood lactate during an incremental test, multiplied with the HR reserve and a weighting factor (Morton et al., 1990). The weighting factor is applied to avoid giving too much emphasis to low-intensity exercise of long duration relative to higher intensities. Alternative HR-based TRIMPs are Edward's TRIMP, which is calculated as a sum of time spent in five HR zones (determined by %HRmax) multiplied by zone-specific weighting factors (Edwards, 1993), and Lucia's TRIMP, in which weighting factors are anchored to ventilatory thresholds instead of arbitrary HR zones (Lucia et al., 2003), although the weighting factors remain somewhat arbitrary. These HR-based TRIMP training load metrics have been a central tool in both research and coaching (Halson, 2014).

Despite some investigations reporting satisfactory performance prediction of the impulse-response model when using the TRIMP concept, serious limitations of TRIMP remain. One of the underlying problems with TRIMP lies within the inherent limitations of using HR as an all-encompassing marker of intensity. The kinetics of the HR response (such as inertia and drift) as well as factors independent of power output (such as environmental conditions and level of arousal) impair its ability to accurately reflect muscular energy requirements, especially in very short, very long, and intermittent (i.e., on/off) efforts. As an example, Vermeire and colleagues analyzed the training of 11 recreational cyclists over 12 weeks and failed to observe a consistent relationship between different TRIMPs and changes in performance (Vermeire et al., 2021). The lack of a consistent dose-response relationship between TRIMP and progression in physical fitness was attributed to limitations with the calculation and the general concept of the training load model used. In a later review article, the same research group discussed limitations of this impulse-response model and concluded that "since the training adaptations performing such different training sessions with a similar [training load] are totally different, the relationship with performance improvement will always be distorted" (Vermeire et al., 2022). HR-derived training load metrics are no longer preferred for quantifying training loads and instead, the use of power meter data to generate a training load metric has become the standard in cycling.

*Popular models for power-derived training load*

With the increasing affordability and validity of portable power meters over the last decade or two, power-based training load metrics have taken over as the most popular way to estimate training load in the field for cycling. Training Stress Score (TSS®; registered trademark by TrainingPeaks) is a training load metric derived from power output recordings (Coggan, 2014). This metric is anchored to the individual functional threshold power (FTP; an estimate of critical power or maximal lactate steady state) and, by definition, 1 hour at this threshold gives 100 TSS. Its calculation uses a concept called normalized power (NP®, also registered trademark by TrainingPeaks), which transforms raw power output values to NP by specific weighting of rolling 30-second average power output. This method aims to account for delayed physiological responses to changes in power output, and for their curvilinear rather than linear nature. TSS for a given training session is calculated as:

$$TSS = \frac{t \cdot NP \cdot IF}{FTP} \cdot \frac{100}{3600s} \tag{1}$$

where IF is the intensity factor and simply the fraction of NP relative to FTP (if NP=FTP, IF becomes 1.0, and 3600 is the number of seconds in an hour. This method has also been transformed to calculate TSS from HR data (for any activity) and from velocity data (for running and swimming) (Thomas, 2017). TSS has been demonstrated to correlate with performance increases equally or better when compared with various HR-derived TRIMPs (Sanders et al., 2017; Vermeire et al., 2021; Wallace et al., 2014).

While TSS might overcome some of the limitations of HR-based methods to quantify training load, it remains largely unvalidated against physiological measurements and has significant limitations that ultimately reduce its ability to predict adaptation. The first limitation, which is common to all other popular training load metrics, is the one-dimensional nature: adaptation is assumed to be independent of intensity. For example, this model assumes that performing all training at an IF of 0.5 would result in the exact same adaptations as performing all training at an IF of 1.1 (but spending ~80% less time doing so), which is not supported by practice or scientific literature.

Another limitation of TSS lies in the independence of duration in the calculation of NP. TSS only considers the distance from FTP when estimating metabolic stress, not the duration for which it is sustained. This is a problem particularly for efforts above FTP (non-steady state exercise). For instance, if we consider a maximal constant-load cycling bout of 20 minutes, the TSS calculated will be the same for minute 2 as it is for minute 19. However, metabolic stress will be close to maximal achievable levels by the end of the task, hence the adaptive stimulus of minute 19 is likely greater than that of minute 2. Indeed, high-intensity interval training relies on this concept; otherwise, there would be no benefit to doing repetitions that generate greater metabolic stress and strain. To overcome this limitation, a power-based training load metric must be able to estimate physiological strain with the help of the entire power-duration curve to assess how close to their maximum rates and capacities each energy system is operating at a given moment. The proposed training load metric in this paper (named strain score, SS), despite also being power-based, considers time spent above the maximal metabolic steady state to estimate the level of strain. The mathematical basis of calculating SS will be introduced in section III.

## III. THREE DIMENSIONS OF TRAINING LOAD

Muscles utilize three complementary energy systems to power muscle contractions. During muscle contraction, even intense contractions, the concentration of ATP stays relatively stable due to the instant replenishment of the ATP that is hydrolyzed mainly by contractile machinery and ion pumping. Resynthesis of ATP is achieved by 1) the alactic anaerobic (PCr/immediate) system; 2) the lactic anaerobic (glycolytic) system; and 3) the aerobic (oxidative) system, and the systems activate in this order, albeit in close succession, at the onset of exercise. While these systems remain active simultaneously with continued contractions, their relative contribution to energy production depends heavily on the duration and intensity of the exercise bout. For a comprehensive review on exercise energy metabolism, see Hargreaves & Spriet (2020).

### *Rationale for energy-system specific adaptation*

Vast amounts of literature exist to demonstrate that training at different ends of the power-duration spectrum elicits different performance adaptations (MacInnis et al., 2022; Viru & Viru, 2000). For example, 8 weeks of continuous running training at moderate intensities elicited improvements in aerobic capacity and running distance in a 30-min test, while short sprint training showed no improvements in these variables but did improve the 50 and 100 m sprint times (Callister et al., 1988). Furthermore, a combination of both training types improved both the aerobic and the sprint outcomes (Callister et al., 1988). Continuous or intermittent training relying heavily on the oxidative system is well-established to promote improvements in maximal aerobic capacity (Hickson et al., 1977) and critical power (CP) (Collins et al., 2022).

In addition, performing all-out sprint interval training, such as 30-second sprints on a cycling ergometer, has been shown to improve both aerobic and anaerobic performance and increase the capacity of all systems (Gibala et al., 2006; Hazell et al., 2010). In this form of exercise, despite the relatively short duration of the work bout, the oxidative system is fully activated at the end of the first sprint and contributes an increasingly greater share of the ATP in subsequent sprints, so that the muscle receives signals to augment all three energy systems (Bogdanis et al., 1996; Parolin et al., 1999). Conversely, sprint training involving only very short sprint bouts (5-10 s) with long recovery has been shown to decrease (Dawson et al., 1998) or lead to no change (Linossier et al., 1993) in markers of oxidative capacity, in line with its low dependence on the aerobic system to provide ATP for the task.

As there is evidence of the three energy systems responding independently with specific adaptations to each, it could be appropriate to model the “fitness” of each system independently rather than using a global amalgamation of them, which inevitably results in compromising their true attributes for simplicity. By quantifying training loads across these three systems, our proposed approach separates each system, resulting in three parallel impulse-response models, each corresponding to an energy system.

*From energy systems to the power-duration relationship*

What are the limiters for instant power output and sustained power output and why should they be considered in performance modeling? Individual endurance exercise capacity at a given intensity can be predicted from a series of maximal efforts in the severe exercise domain (Poole et al., 2016). The power-duration relationship for cycling is generally accepted to be hyperbolic, and its asymptote is critical power (CP) (Hill, 1993). The CP model was originally proposed for single muscle group exercise (Monod & Scherrer, 1965) but was later adapted to describe power-duration behaviour in whole body exercise (Hughson et al., 1984; Moritani et al., 1981).

CP has been suggested as the gold standard delineating the heavy intensity domain from the severe intensity domain (Jones et al., 2019). As such, CP is the ceiling for the rate of sustainable oxidative energy provision, as it separates exercise intensities for which a metabolic steady state (i.e., stable oxygen uptake) is achievable from intensities for which it is not (Black et al., 2017; Poole et al., 1988). According to the standard interpretation of the CP model, exercising above CP results in the gradual depletion of W′—the finite work capacity above CP—until task failure occurs, whereas exercising below CP does not draw on W′, implying (mathematically) that exercise below CP can be performed indefinitely (Poole et al., 2016). Despite its popularity, this 2-parameter model incorrectly assumes that when duration approaches zero, power output approaches infinity. A 3-parameter (3-CP) power duration-model includes maximal sprint power (Pmax) as a third parameter, addressing the unrealistic assumption of unlimited W′ utilization rates (Morton, 1996).

The equation defining the popular 2-parameter power-duration relationship is:

$$t_{lim} = \frac{W'}{P - CP}$$

(2)

This CP model assumes an unlimited rate of W′ expenditure. A 3-parameter (3-CP) power duration-model that includes Pmax as a third parameter has been shown to result in better fit for short exercise durations (Morton, 1996; Vinetti et al., 2019). It is defined as:

$$t_{lim} = \frac{W'}{P - CP} - \frac{W'}{Pmax - CP}$$

(3)

where Pmax represents the highest instant power that can be produced for a very short duration and is reached at an intensity at which the alactic/PCr system is providing most the ATP.

A consequence of the 3-parameter model is that the theoretical Pmax is only achievable in a fatigue-free state (Morton, 1990, 2006). The highest possible power output, maximum power available (MPA), is a function of the remaining W′ and decreases with W′ expenditure (W′exp). Therefore, MPA changes according to the equation:

$$MPA = Pmax - (Pmax - CP) \cdot \frac{W'exp}{W'}$$

(4)

where W′exp = W′ - W′bal and W′bal is the amount of W′ remaining at a given moment. For power outputs P > CP, task failure occurs when P = MPA, not when W′ is fully depleted.

The three parameters (CP, W′, and Pmax) and the power-duration curve they define can be interpreted to contain information of the three energy systems: their maximal rates and their capacities (for the anaerobic systems, since the aerobic system is not considered capacity-limited) (Morton, 2006). Consequently, a 3-parameter CP model may be used to estimate the contribution of each energy system to the power output at a given moment in time.

The connection between the three parameters of the power-duration relationship and the energy systems is illustrated in Figure 2 (A, B).

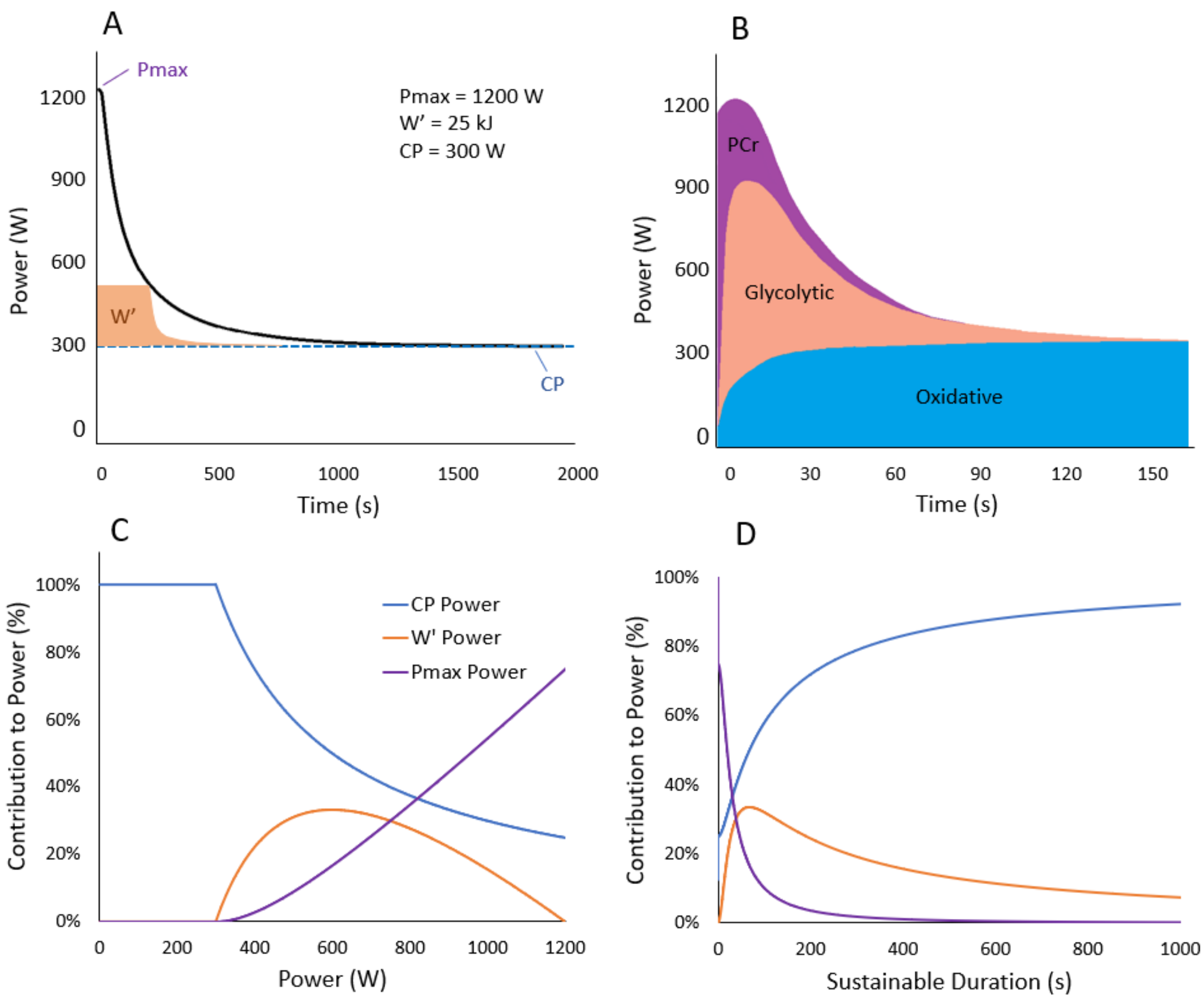

Figure 2. Linking the three energy systems with the three parameters of the power duration relationship. A) The power output attainable for a given duration can be predicted from a 3-parameter critical power (CP) model, in which maximal power output is limited by Pmax, highest aerobically sustainable power output is limited by CP, and the finite work capacity above CP, curvature constant (W′), is primary limited by buffering capacity and muscle energy stores available for substrate-level phosphorylation. B) A schematic representation depicting the contribution of each energy system to power production in all-out exercise of 2.5 minutes. The relative contribution of each system is related to the parameters $P_{max}$ (the PCr system), W′ (the glycolytic system), and CP (the oxidative system). Constructed based on data from Baker et al., (2010), Morton (2006), and Vanhatalo et al., (2007). C) The contribution of the parameters of the 3-parameter CP model to a given power output plotted against power. D) The contribution of the parameters of the 3-parameter CP model to a given power output plotted against the maximal sustainable constant-load duration (D).

*Quantification of energy system contributions*

Continuous power data may be converted to energy system-specific work based on power output and how close the individual is to their MPA at a given moment in time. Furthermore, these values may be used as input values for training load imposed on each energy system. The theoretical basis of this conversion is described below.

At a given moment of time, the power output (*P*) is a sum of energy provision by each energy system. For example, the oxidative system (represented by CP) is unable to provide enough ATP to power 600 W; therefore, some of the energy must come from the alactic system (Pmax) and the lactic system (W′). How much energy comes from each system for each athlete? When an athlete's CP and Pmax are known, we may mathematically derive the contribution of each system to a given *P*. Using the MPA formula (Equation 4), we compute a power duration relationship for a constant power performed to failure in the special case where *P* = *MPA*:

$$P = Pmax - (Pmax - CP) \cdot \frac{(P - CP)t}{W'}$$

(5)

This is the constant power sustainable for time *t*. Solving for *P*:

$$P = CP + \frac{(Pmax - CP)W'}{W' + (Pmax - CP)t}$$

(6)

We now have the power-duration formula for the 3-parameter CP model. We can rearrange this formula to compute the contributions of CP, Pmax and W′ to power P, as a function of CP, W′, Pmax and P by separating into 3 separate terms:

$$P = CP + \frac{(P - CP)^2}{Pmax - CP} + \frac{(P - CP)^2 t}{W'}$$

(7)

From this basis, the contributions of CP, Pmax and W′ may be derived, assuming *CP* represents the maximum rate of aerobic power. Therefore:

Contribution of *CP*:

$$If\ P < CP, P_{CP} = P; If\ P > CP, P_{CP} = CP \tag{8}$$

Contribution of *Pmax*:

$$P_{Pmax} = \frac{(P-CP)^2}{Pmax-CP}, \text{where } P > CP \tag{9}$$

Contribution of *W′:*

$$P_{W\prime} = \frac{(P-CP)^2}{W\prime} = P - CP - P_{Pmax}, \text{where } P > CP \tag{10}$$

Equation (9) can be understood through the following reasoning. When exercising at CP, Pmax (alactic system) has no contribution to power, since the aerobic system can fully account for the ATP turnover. In contrast, when exercising at Pmax, Pmax has 100% contribution to the power (above CP), as instant maximal power is not limited by W′. Therefore, between CP and Pmax, the contribution of Pmax must get a value between 0 and 1. Equation (10) represents what must be left for W′ (lactic system) to contribute after subtracting the effects of CP's and Pmax's contributions. Finally, Equation (8) can be understood as the aerobic system contributing fully to all power up to CP, but no more to any increases in power above it.

To demonstrate the calculation: if the athlete with a CP of 300 W and Pmax of 1200 W is sprinting at Pmax, then all of the power generated that is above CP is attributed to Pmax. Any power output between CP and Pmax will have some contribution from all three parameters. If they are exercising at 1000 W, according to equations (8) to (10), 544 W is attributed to Pmax, 156 W to W′, and 300 W to CP. If exercising at 400 W, then the shares of power production from the three systems become 11 W, 89 W, and 300 W, respectively. If cycling at CP or below, 100% of power output is attributed to CP. The relative contributions of the parameters CP, W′, and Pmax, calculated using equations (8), (9), and (10), are illustrated in Figure 2 as plotted against power (2C) and against sustainable duration (2D).

How can these power allocations be translated into a training load input reflecting the share of each parameter accurately? As argued above, training-induced strain is a function of not only intensity, but also duration. Here, the concept of MPA becomes important, because it allows us to assess how close to their limit the individual is exercising at a given moment of time. When cycling at power output P (>CP) for long enough to deplete most of W′, MPA will approach P, making exercise increasingly difficult and elevating the strain induced by each second of additional exercise. We may estimate the strain coefficient ($k_{strain}$, unitless) by:

$$k_{strain} = \frac{Pmax - MPA + CP}{Pmax - P + CP} \tag{11}$$

Strain rate (SR) is calculated by multiplying $k_{strain}$ by power output (*P*):

$$SR = k_{strain} \cdot P \tag{12}$$

The SR (units: W) is calculated on a second-by-second basis for continuously recorded power data.

Finally, strain score (SS) quantifies how much total strain an athlete endures during an activity. To allow for comparison to TSS with values of similar magnitude, second-by-second SR is normalized to CP so that one-hour (3600 s) at CP equates to 100 SS.

*SS* is calculated as:

$$SS = \sum SR \cdot \left(\frac{Pmax}{CP^2} \cdot \frac{100}{3600s}\right) \tag{13}$$

The factor $Pmax / CP^2$ is used to achieve 100 *SS*/h when exercising at *CP*, as can be demonstrated by substituting the equations for *SR* (12) and $k_{strain}$ (11) into Equation (13).

This SS can be subdivided into $SS_{Pmax}$, $SS_{W'}$, and $SS_{CP}$ by using the system-specific strain rates $SR_{CP}$, $SR_{W'}$, and $SR_{Pmax}$, calculated from the respective power contributions $P_{Pmax}$, $P_{W'}$, and $P_{CP}$ (Equations 8-10).

For example, if exercising fresh (i.e., W′exp = 0; see Equation (4)), and power P equals CP, MPA = Pmax and the $k_{strain}$ gets a value of 0.27. However, when available W′ is fully exhausted (W′exp = W′), MPA = CP and $k_{strain}$ becomes 1.0. For this example athlete, the model implies that exercising at CP is approximately four times more strenuous when W′ has been depleted compared with when they were fresh. The impact of falling MPA on strain is illustrated in Figure 3.

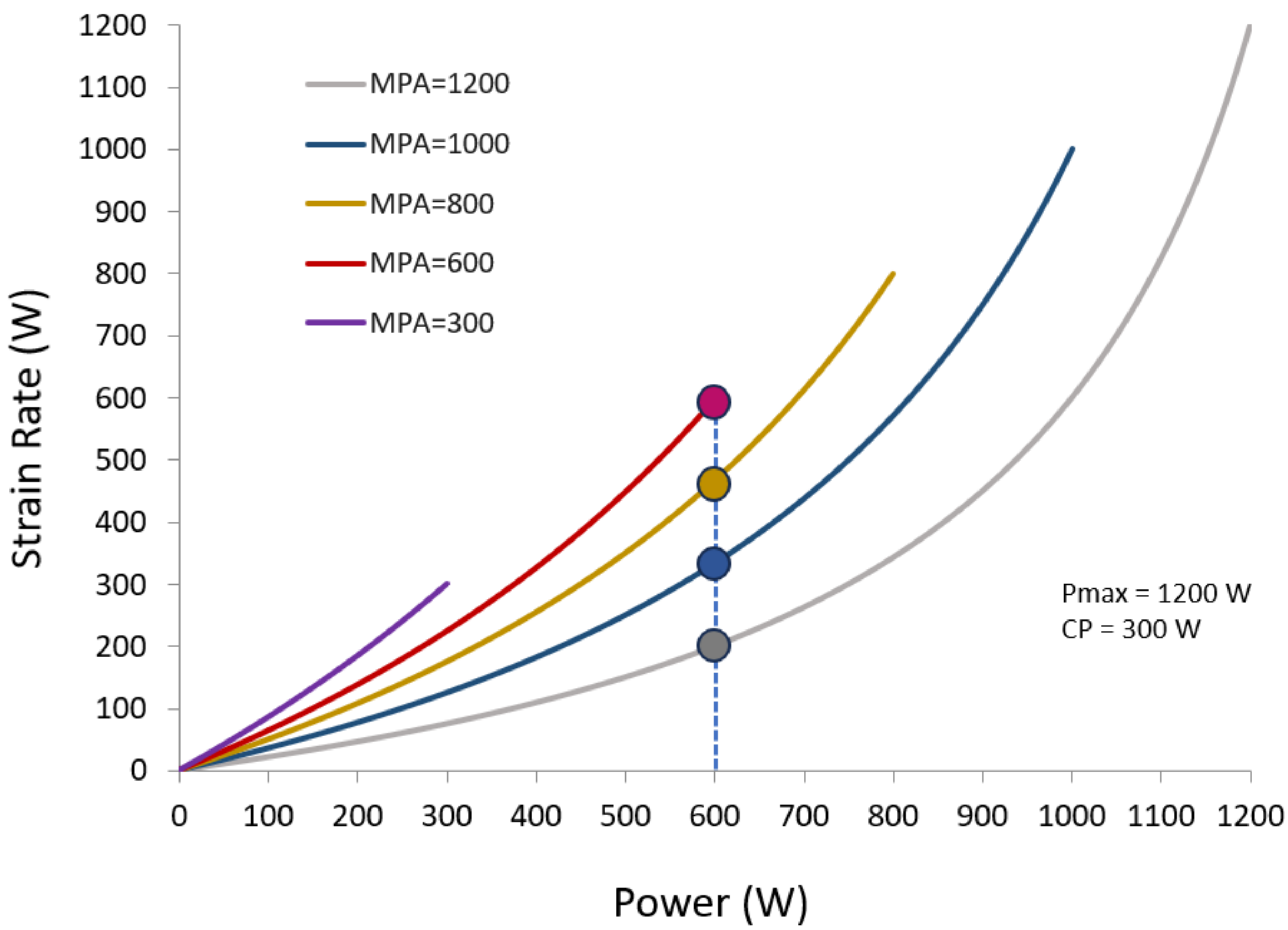

Figure 3. Maximum power available (MPA) influences the strain experienced at a given power output. MPA is a function of the amount of W′ expended. To generate 600 W when fresh (MPA

= Pmax = 1200 W), is less strenuous (grey dot) as compared to a situation where MPA is reduced to 1000, 800, or 600 W (blue, yellow, and red dot, respectively). When MPA = CP = 300 W, W′ is fully depleted and generating 600 W is not possible.

Table 1. Example of the strain score calculation for different levels of MPA

| | MPA = 1200 | MPA = 1000 | MPA = 800 | MPA = 600 | MPA = 400 |
|---|---|---|---|---|---|
| **Power output (W)** | | | **$k_{strain}$** | | |
| **100** | 0.21 | 0.36 | 0.50 | 0.64 | 0.79 |
| **200** | 0.23 | 0.38 | 0.54 | 0.69 | 0.85 |
| **300** | 0.25 | 0.42 | 0.58 | 0.75 | 0.92 |
| **400** | 0.27 | 0.45 | 0.64 | 0.82 | 1.00 |
| **800** | 0.43 | 0.71 | 1.00 | | |
| **1200** | 1.00 | | | | |
| **Power output (W)** | | | **SR (W)** | | |
| **100** | 21 | 36 | 50 | 64 | 79 |
| **200** | 46 | 77 | 108 | 138 | 169 |
| **300** | 75 | 125 | 175 | 225 | 275 |
| **400** | 109 | 182 | 255 | 327 | 400 |
| **800** | 343 | 571 | 800 | | |
| **1200** | 1200 | | | | |
| **Power output (W)** | | | **SS ($\cdot s^{-1}$)** | | |
| **100** | 0.008 | 0.013 | 0.019 | 0.024 | 0.029 |
| **200** | 0.017 | 0.028 | 0.040 | 0.051 | 0.063 |
| **300** | 0.028 | 0.046 | 0.065 | 0.083 | 0.102 |
| **400** | 0.040 | 0.067 | 0.094 | 0.121 | 0.148 |
| **800** | 0.127 | 0.212 | 0.296 | | |
| **1200** | 0.444 | | | | |
| **Power output (W)** | | | **SS ($\cdot h^{-1}$)** | | |
| **100** | 29 | 48 | 67 | 86 | 105 |
| **200** | 62 | 103 | 144 | 185 | 226 |
| **300** | 100 | 167 | 233 | 300 | 367 |
| **400** | 145 | 242 | 339 | 436 | 533 |
| **800** | 457 | 762 | 1067 | | |
| **1200** | 1600 | | | | |

Note: Calculated for an athlete whose CP = 300 W and Pmax = 1200 W

MPA = Maximum power available; $k_{strain}$ = strain coefficient, SR = strain rate; SS = strain score.

*Advantages of the novel strain score*

What practical advantages does the novel training load metric SS provide over existing metrics such as total work or TSS? First, since it considers duration of exercise above CP in addition to the distance from CP, it may more accurately estimate the metabolic stress, and consequently strain, than other power-based metrics.

A comparison against TSS and work completed is illustrated in Figure 4.

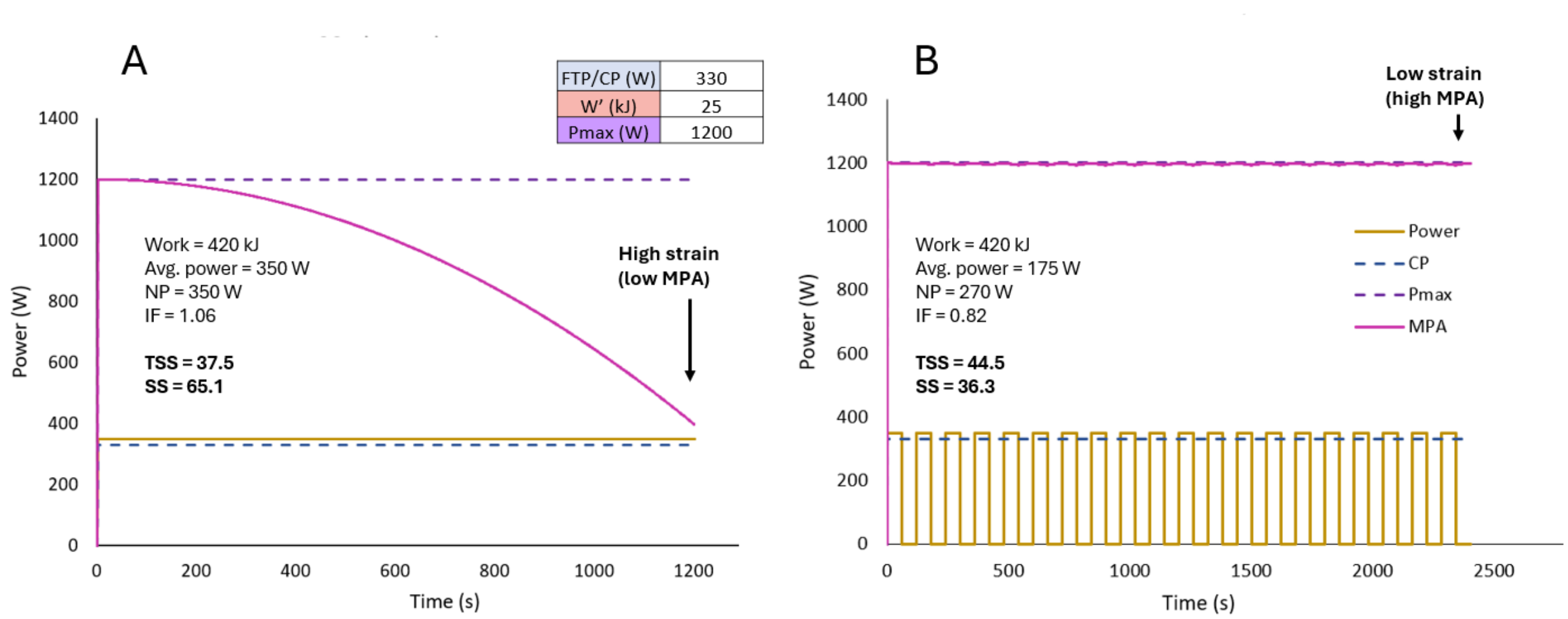

Figure 4. Limitations of using work completed or training stress score (TSS) to estimate the load of a training session. *A*) A 20-min continuous effort at 350 W and associated training load metrics. *B*) A 20x1-min interval session at 350 W with 1-min passive rest. The example athlete's parameters are listed in panel A. In both *A* and *B*, work completed is the same. TSS for scenario *B* is higher than for *A* despite greater expected metabolic perturbations in *A*. Strain score (SS) is higher for scenario *A* than for *B*, better reflecting the expected physiological strain of these efforts. At task failure, MPA (maximum power available) equals task power output. FTP = functional threshold power; CP = critical power; W′ = work prime; Pmax = maximal power output; NP = normalized power; IF = intensity factor. FTP and CP are assumed to be equal.

Second, the SS metric introduced has three dimensions, allowing for a more specific quantification of training loads imposed on each energy system. This feature helps overcome the well-recognized limitations of relying on one universal training load metric to encompass all training sessions of different types, and the resulting inaccurate predictions when using an impulse-response model. Nuances of training are not captured if using metrics such as TRIMP, TSS, or total work since they reduce vastly different training responses to a single value. Figure 5 illustrates the effect of quantifying training load using $SS_{CP}$, $SS_{W'}$, and $SS_{Pmax}$ universal, one-dimensional metrics. Note that in Figures 4 and 5, the MPA shown follows the modified 3-parameter model where the W′exp/ W′ factor (see Eq 4) is raised to the 2$^{nd}$ power (Kontro et al., 2024) and the recovery of W′ balance follows the differential model by Skiba et al (Skiba et al., 2015) as the combination of these models have shown to effectively predict intermittent exercise in practice (unpublished data).

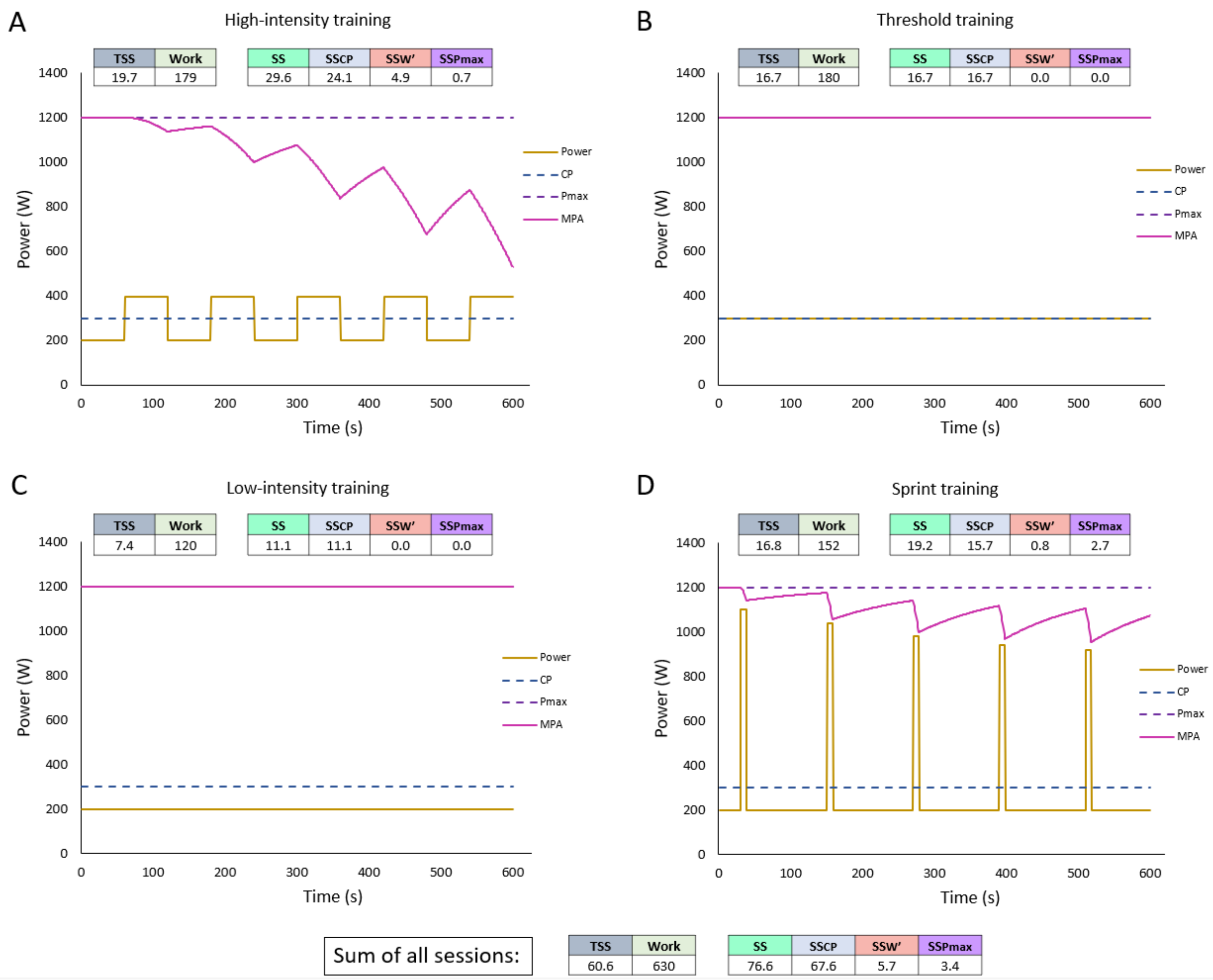

Figure 5. A comparison of total work, training stress score (TSS), and strain score (SS) in four different cycling sessions of 10 minutes. Despite the different types of training sessions, the TSS values resulting for A, B, and D are similar. $SS_{CP}$, $SS_{W'}$, and $SS_{Pmax}$ represent the strain on the aerobic, glycolytic, and PCr energy systems, respectively, which reflects the training load of critical power (CP), work prime (W′), and maximal power (Pmax), respectively. The breakdown of SS into three dimensions allows for a more detailed analysis of the type of training performed. MPA = maximum power available, CP = critical power; W′ = work prime; Pmax = maximal power output.

## IV. THE THREE-DIMENSIONAL IMPULSE-RESPONSE MODEL

The three training load metrics described in the previous section can separately be used to model performance across intensities and durations. The rationale to do this relies on the well-established properties of the power-duration relationship of exercise (Jones & Vanhatalo, 2017; Morton, 2006), as well as the principle of specificity, which is explained by the tendency of adaptation to only occur in the systems stressed and not others.

First, a brief introduction to the standard, one-dimensional impulse-response model is given. Consider the equation for performance by Morton et al. (1990), who used this model to predict running performance:

$$p(t) = k_1 g(t) - k_2 h(t) \tag{14}$$

Where $p$ is performance response, $t$ is time (in days), $k_1$ and $k_2$ are weighting factors for fitness and fatigue, respectively, $g(t)$ is fitness, and $h(t)$ is fatigue. The values of $g(t)$ and $h(t)$ depend on the stimulus (i.e., training load), denoted below as $w(t)$, and the time interval, $i$, at which the stimulus is repeated:

$$g(t) = g(t-i)e^{-i/\tau_1} + w(t) \tag{15}$$

$$h(t) = h(t-i)e^{-i/\tau_2} + w(t) \tag{16}$$

The time constants $\tau_1$ and $\tau_2$ define the kinetics of fitness and fatigue progression, respectively. An optimal taper for peak performance involves manipulating the training load so that the loss in fitness is minimized while the loss in fatigue is maximized (the effect of varying training load on predicted performance is illustrated in Figure 6).

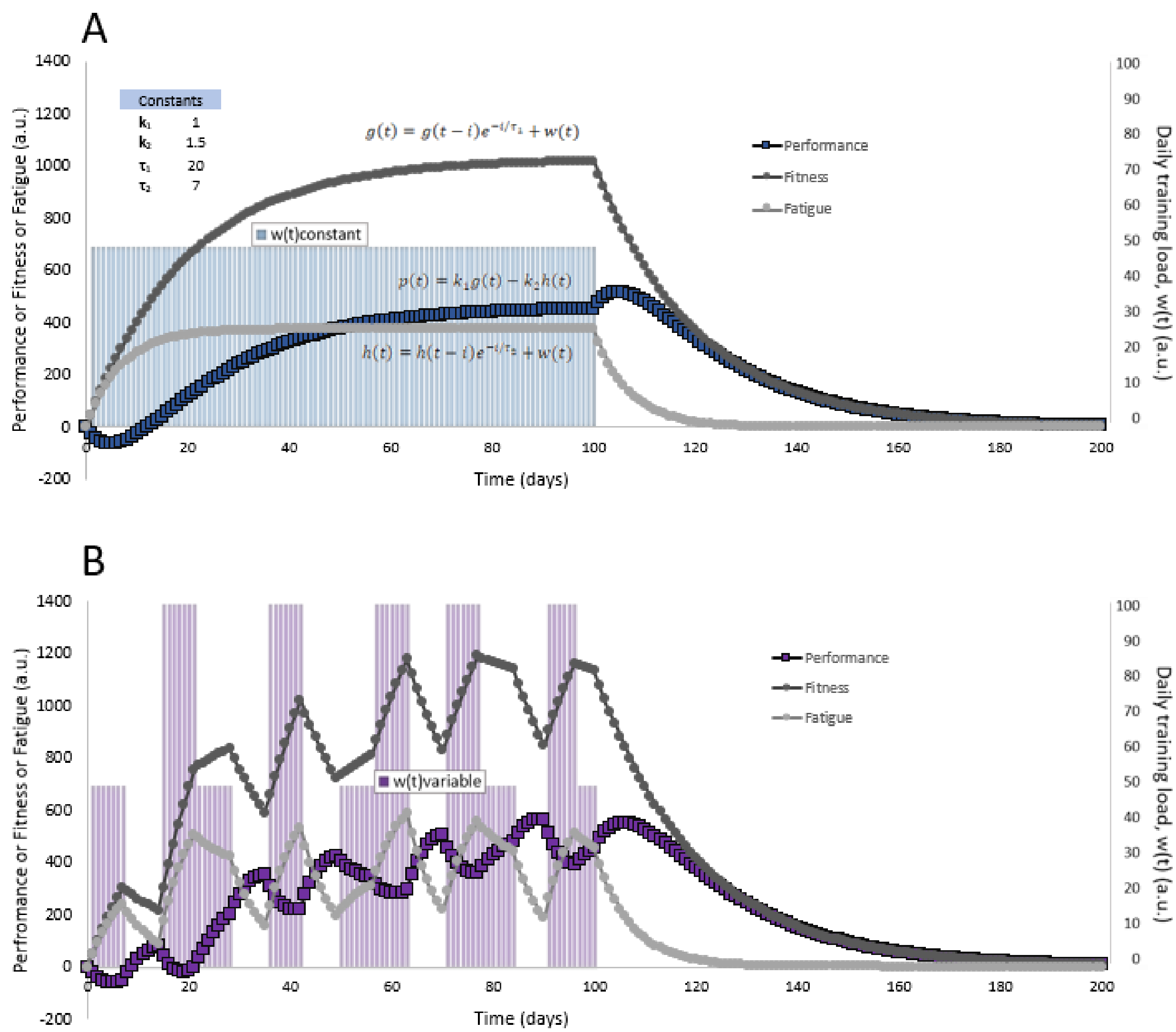

Figure 6. Example of predicted progression of fitness (g), fatigue (h), and performance (p) during 100 days of daily training at a constant (A) or variable (B) average training load of w(t) = 50 followed by 100 days of detraining at w(t) = 0, modelled using equations 14, 15, and 16. Values for the weighting factors $k_1$ and $k_2$ are 1 and 1.5, respectively, and the time constants $\tau_1$ and $\tau_2$ are 20 days and 7 days, respectively. Units for g, h, and p on the y-axis are arbitrary. Despite an overall identical accumulated training load of 5000 units over the training period, the performance peak in the variable program (B) is greater than that of the constant program (A).

Popular training platforms use a slight variation of Banister's original formula, where fitness is calculated as an exponentially weighted moving average of daily training load:

$$g(t) = g(t-1)e^{-1/\tau_1} + w(t)(1 - e^{-1/\tau_1})$$

(17)

Or, using popularized terminology:

$$CTL = CTL_{yesterday}e^{-1/\tau_1} + TSS_{today}(1 - e^{-1/\tau_1})$$

(18)

Where CTL stands for chronic training load and is equivalent to fitness.

With varying success, the impulse-response model has been used to relate training load to changes in performance in different sports, including swimming (Calvert et al., 1976; Hellard et al., 2005; Mujika et al., 1996), triathlon (Millet et al., 2002), cycling (Busso et al., 2002; Busso et al., 1991; Vermeire et al., 2021), and running (Morton et al., 1990; Wallace et al., 2014; Wood et al., 2005). However, substantial uncertainty remains around the model's ability to predict performance, which might be in part explained by the difficulty of quantifying training load as outlined in the prior sections.

*Implementation of the 3D-impulse-response model*

As detailed in section III, the new training load metric SS makes use of continuous power meter data to convert second-by-second work to strain based on an athlete's proximity to their MPA (Baron Biosystems Ltd., 2024; Kontro et al., 2024). Second-by-second strain for each fitness parameter can be calculated by multiplying strain rate with the ratio corresponding to the contribution of each energy system (Table 1).

When calculated in this manner, training load (w(t)) is divided into three components and fed into three parallel impulse-response models. This way, the three-dimensional model produces three separate performance management charts (PMCs) instead of one. Figure 7 shows a simulation of how fitness, fatigue, and performance of each system could evolve in response to a daily load of w(t) = 80, w(t) = 18, and w(t) = 2 (A.U.) for the oxidative (CP), glycolytic (W′), and PCr (Pmax) systems, respectively.

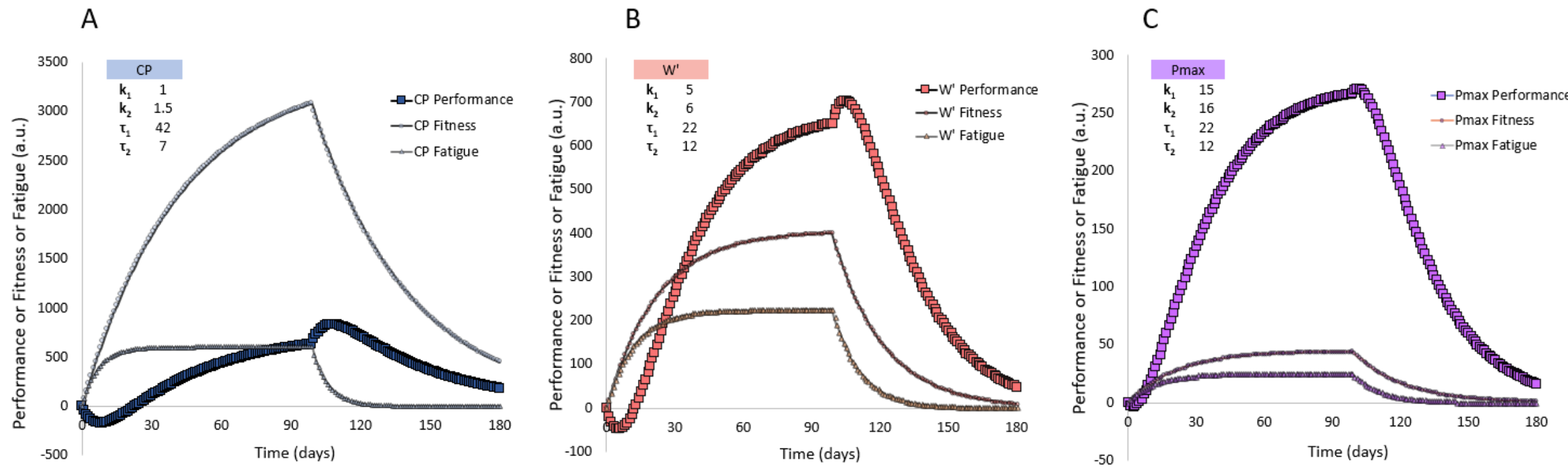

Figure 7. A hypothetical example of how a three-dimensional impulse model would predict the changes in performance for the oxidative system (A), the glycolytic system (B), and the PCr system (C) based on changes in fitness and fatigue for each system. The performance outcome (a.u.) at a given timepoint can be translated into critical power (CP), work prime (W′), and maximal power output (Pmax) using a conversion factor to get units of W, kJ, and W, respectively. Parameter values for the model (weighting factors $k_1$ and $k_2$ and the time constants $\tau_1$ and $\tau_2$) can be experimentally determined for each system. In this simulation, w(t) is repeated daily and is 80, 18, and 2, for CP, W', and Pmax, respectively.

Translation of the arbitrary units for fitness to directly informative units of power and work is preferable, as these fitness markers represent the parameters of the 3-parameter CP model. An example of real training data when visualized either in a traditional performance-management chart (PMC) or the three-dimensional model is shown in Figure 8. Notably, the three parameters CP, W′, and Pmax, do not always evolve in the same direction with increases or decreases in overall training load, but they respond in distinct ways depending on the type of training performed.

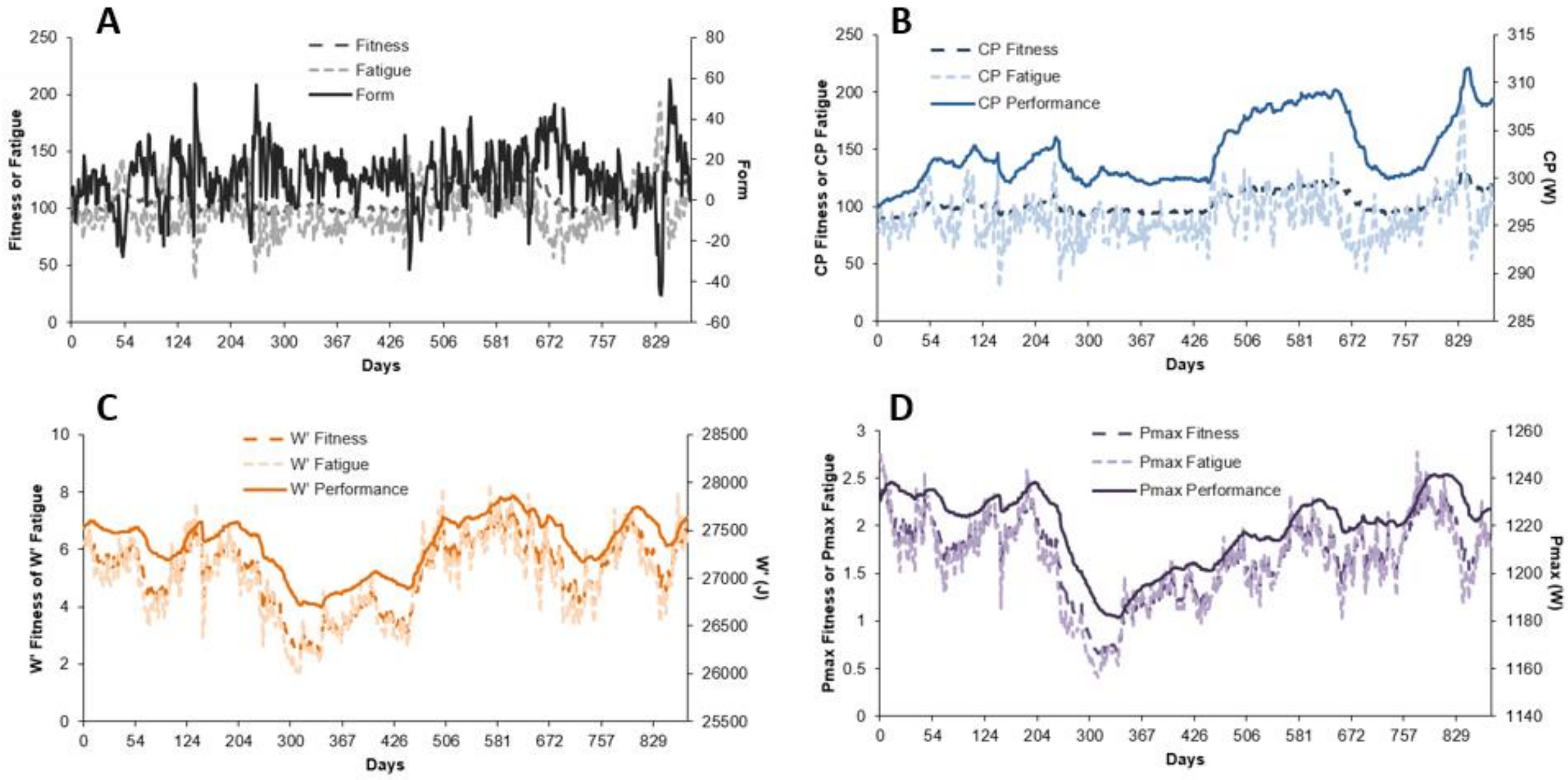

Figure 8. An example of an athlete's training data collected over ~2.5 years visualized in a performance management chart (PMC). A) One-dimensional PMC with overall fitness, fatigue, and form (performance readiness). Panels B, C, and D illustrate the three-dimensional model and the independent evolution of critical power, W′, and Pmax, respectively, over the same period.

*What are the model parameters for each energy system?*

It is expected that the parameters (weighting factors, k, and time constants, τ) for the adaptation of each energy system are different. While no published data exist to support the energy-system specific model parameters, the assumption of their independence is suggested by data demonstrating that anaerobic performance can be enhanced in shorter time frames (different τ) and with much smaller doses (different k) than aerobic performance (Gibala et al., 2006; Mølmen et al., 2025; Roberts et al., 1982; Ross & Leveritt, 2001). This is well-established in practice, which is why effective training plans have historically used a periodization strategy of more high-volume, low-intensity training early in the season and increasing amounts of high-intensity training closer to competition (Gilmour & Lydiard, 2000; Hawley, 1993). A recent meta-analysis suggests that while improvements in aerobic capacity can be achieved with sprint intervals, high-intensity intervals, or continuous exercise, the time-course of change is proportional to intensity (i.e., slower for lower intensities) (Mølmen et al., 2025). In addition, the gains in $VO_{2max}$ are greater with high-intensity intervals and continuous exercise training than with sprint interval training, which is consistent with their greater accumulated stress on the aerobic system (Mølmen et al., 2025). This evidence suggests that emphasizing higher intensities (as opposed to moderate / threshold intensities) closer to the competition date is beneficial for maximizing high-intensity performance rapidly, but the aerobic system may be trained over longer time frames.

The weighting factors (k) and time constants (τ) in the literature, both for fitness and fatigue, are based on a one-dimensional impulse-response model. Therefore, the reported parameters are reflecting some type of average of the performance of the responses of different energy systems. This might not be of great significance in sports where one energy system is clearly dominant and almost exclusively responsible for the observed changes in performance (e.g., the oxidative system in marathon running, or the alactic anaerobic system in weightlifting), and most training is also accumulated in the same intensity domain. However, countless sports (e.g., track cycling, track running, swimming, rowing, speed skating) involve maximal efforts spanning a range of less than one minute to several minutes and/or are highly stochastic in power demands. In addition, sports like team sports or road cycling combine long durations of low-intensity exercise with intermittent high-intensity efforts. Therefore, in most sports, successful athletes do not rely only on their CP, as W' and Pmax substantially contribute to their competitive success. In these sports, modeling fitness and fatigue of the energy systems separately is expected to produce a more complete picture of athletic capabilities spanning a wide range of durations and intermittent exercise performance.

The model parameters are individualized to the athletes as training responsiveness shows large inter-individual variation (Bonafiglia et al., 2016; Bouchard & Rankinen, 2001). This variation has been suggested to arise from variation in factors including genetics (Bouchard et al., 2011), diet (Hawley et al., 2006), indicators of stress (Nummela et al., 2010), initial fitness level (Thomas et al., 1985), and generic training prescription leading to unequal internal load between individuals (Weatherwax et al., 2019). Ideally, the values for τ and k are established individually for each athlete and recalibrated periodically. In practice, training platforms use fixed parameters and the τ parameter is often estimated to be 42 days for fitness and 7 days for fatigue. Values for k cannot be compared across models using different training load inputs (which are in arbitrary units not easily transformed to one another), but time constants are unlikely to depend on the training load metric. Table 2 summarizes some time constants reported in the literature.

Table 2. Reported time constants for fitness (**$\tau_1$**) and fatigue (**$\tau_2$**) in endurance sports

| **Study** | **$\tau_1$** | **$\tau_2$** |
|---|---|---|
| Busso et al. (2003) (Period 1) | 41 ± 15 | 9 ± 6 |
| Busso et al. (2003) (Period 2) | 35 ± 12 | 13 ± 3 |
| Chalencon et al. (2015) (Banister model) | 51 ± 14 | 10 ± 9 |
| Chalencon et al. (2015) (Variable dose-response model) | 48 ± 16 | 8 ± 4 |
| Morton et al. (1990) (Subject 1) | 50 | 11 |
| Morton et al. (1990) (Subject 2) | 40 | 11 |

## V. LIMITATIONS OF THE MODEL

While addressing some limitations inherent to the simple Banister model, the three-dimensional model described above is not free from limitations. First of all, several assumptions, some of which are acknowledged to be unrealistic, must be made to keep the model sufficiently simple to be useful. The underlying assumptions are as follows:

1) Each energy system adapts in proportion to its contribution to energy turnover, amplified by the magnitude of strain.
2) The adaptive process of an energy system can be uniformly described by simple parameters.
3) The fitness weighing factor (constant $k_1$) does not change over time.
4) The contribution of each energy system can be derived from a three-parameter critical power model.
5) At the onset of exercise, energy provided by the aerobic system is immediately available at a maximum rate.
6) The energy supply by each energy system remains constant at a given power output.
7) There is no change in efficiency over the course of an exercise bout.
8) The power-duration parameters CP, W′, and Pmax do not change over the course of exercise.

Of the listed assumptions, number 1 is currently unknown, as the presented model has not been under strict scientific scrutiny. Assumptions 2 and 3 are common to all Banister models. Number 2 is an inherent quality of a systems model, which a reductionist approach could argue against. For example, both mitochondrial and cardiac adaptations may contribute to the fitness of the aerobic system, but their time course of change is very different despite the model using a single time constant (Arbab-Zadeh et al., 2014; Egan et al., 2013). Number 3 is unknown but possibly false, as there likely is some maximum to physiological qualities that cannot respond further by infinite increases in the stimulus – but whether this maximum is often approached in practice is uncertain. Number 4 has also not been experimentally confirmed despite a solid theoretical basis; however, it is acknowledged that assumptions 5-7 skew the calculation of energy system-specific contributions if not corrected (not implemented in this model). Numbers 5 and 6 are false based on the well-described kinetics of the $\dot{V}O_2$ response and the delayed contribution of the glycolytic system relative to the PCr system (Davies et al., 2017). Despite this fact, the oxygen deficit created at the onset of exercise—and erroneously ascribed to the aerobic system in the model—will eventually be repaid by the aerobic system; therefore, this simplification may not reduce the accuracy considerably. Assumption number 8 is also inaccurate in the light of emerging evidence, and the parameter most affected might be W', which is sensitive to glycogen availability (Clark et al., 2019).

Despite its limitations, the model presented is an attempt to synthesize current understanding of adaptive responses and the principle of specificity with a systems model. While we believe it to be an improvement to existing models, we also acknowledge the lack of experimental data to assess its performance. The model predicts future "fitness signatures" (i.e., the parameters CP, W', and Pmax), which means predicting points of failure during maximal exercise. To experimentally validate the model, as done by previous researchers with other models, a performance prediction from a carefully quantified training block needs to be compared with performance outcomes. Given the popularity of power meters, the ability to analyze big datasets from real-life training may prove helpful in fulfilling this goal.

## VI. CONCLUSIONS

The three-dimensional impulse-response model presented in this paper has some improvements compared to the simple Banister model. A one-dimensional model is widely used in endurance sports—particularly in cycling—despite its acknowledged shortcomings. By quantifying training load as three separate input metrics in the model, each reflecting the fitness of their respective energy system, the training principle of specificity is accounted for, theoretically resulting in more precise performance predictions. In addition, the quantification of training loads using the MPA concept, derived from a three-parameter power-duration model, appropriately considers exercise duration and not intensity alone as a variable affecting the level of metabolic stress during non-steady state exercise, thus adjusting the accumulated training load used for the input value in the model.

VII. REFERENCES

Arbab-Zadeh, A., Perhonen, M., Howden, E., Peshock, R. M., Zhang, R., Adams-Huet, B.,…Levine, B. D. (2014). Cardiac remodeling in response to 1 year of intensive endurance training. *Circulation*, *130*(24), 2152-2161. https://doi.org/10.1161/circulationaha.114.010775

Baker, J. S., McCormick, M. C., & Robergs, R. A. (2010). Interaction among Skeletal Muscle Metabolic Energy Systems during Intense Exercise. *J Nutr Metab*, *2010*, 905612. https://doi.org/10.1155/2010/905612

Banister, E. W., & Calvert, T. W. (1980). Planning for future performance: implications for long term training. *Can J Appl Sport Sci*, *5*(3), 170-176.

Banister, E. W., Calvert, T. W., Savage, M. V., & Bach, T. (1975). A systems model of training for athletic performance. *Australian Journal of Sports Medicine*, *7:57-61*.

Baron Biosystems Ltd., 2024 (webpage). MPA™ – Maximal Power Available https://baronbiosys.com/maximal-power-available/ Accessed Feb 2025.

Black, M. I., Jones, A. M., Blackwell, J. R., Bailey, S. J., Wylie, L. J., McDonagh, S. T.,…Vanhatalo, A. (2017). Muscle metabolic and neuromuscular determinants of fatigue during cycling in different exercise intensity domains. *J Appl Physiol (1985)*, *122*(3), 446-459. https://doi.org/10.1152/japplphysiol.00942.2016

Bogdanis, G. C., Nevill, M. E., Boobis, L. H., & Lakomy, H. K. (1996). Contribution of phosphocreatine and aerobic metabolism to energy supply during repeated sprint exercise. *J Appl Physiol (1985)*, *80*(3), 876-884. https://doi.org/10.1152/jappl.1996.80.3.876

Bonafiglia, J. T., Rotundo, M. P., Whittall, J. P., Scribbans, T. D., Graham, R. B., & Gurd, B. J. (2016). Inter-Individual Variability in the Adaptive Responses to Endurance and Sprint Interval Training: A Randomized Crossover Study. *PLoS One*, *11*(12), e0167790. https://doi.org/10.1371/journal.pone.0167790

Bouchard, C., & Rankinen, T. (2001). Individual differences in response to regular physical activity. *Med Sci Sports Exerc*, *33*(6 Suppl), S446-451; discussion S452-443. https://doi.org/10.1097/00005768-200106001-00013

Bouchard, C., Sarzynski, M. A., Rice, T. K., Kraus, W. E., Church, T. S., Sung, Y. J.,…Rankinen, T. (2011). Genomic predictors of the maximal $O_2$ uptake response to standardized exercise training programs. *J Appl Physiol (1985)*, *110*(5), 1160-1170. https://doi.org/10.1152/japplphysiol.00973.2010

Burnley, M., Bearden, S. E., & Jones, A. M. (2022). Polarized Training Is Not Optimal for Endurance Athletes. *Med Sci Sports Exerc*, *54*(6), 1032-1034. https://doi.org/10.1249/mss.0000000000002869

Busso, T., Benoit, H., Bonnefoy, R., Feasson, L., & Lacour, J. R. (2002). Effects of training frequency on the dynamics of performance response to a single training bout. *J Appl Physiol (1985)*, *92*(2), 572-580. https://doi.org/10.1152/japplphysiol.00429.2001

Busso, T., Carasso, C., & Lacour, J. R. (1991). Adequacy of a systems structure in the modeling of training effects on performance. *J Appl Physiol (1985)*, *71*(5), 2044-2049. https://doi.org/10.1152/jappl.1991.71.5.2044

Callister, R., Shealy, M. J., Fleck, S. J., & Dudley, G. A. (1988). Performance Adaptations to Sprint, Endurance and Both Modes of Training. *The Journal of Strength & Conditioning Research*, *2*(3), 46-51.
Calvert, T. W., Banister, E. W., Savage, M. V., & Bach, T. (1976). A Systems Model of the Effects of Training on Physical Performance. *IEEE Transactions on Systems, Man, and Cybernetics*, *SMC-6*(2), 94-102. https://doi.org/10.1109/TSMC.1976.5409179
Clark, I. E., Vanhatalo, A., Thompson, C., Wylie, L. J., Bailey, S. J., Kirby, B. S.,…Jones, A. M. (2019). Changes in the power-duration relationship following prolonged exercise: estimation using conventional and all-out protocols and relationship with muscle glycogen. *Am J Physiol Regul Integr Comp Physiol*, *317*(1), R59-r67. https://doi.org/10.1152/ajpregu.00031.2019
Coggan, A. (2014). *What is TSS? By TrainingPeaks*. Retrieved 24/11/2020 from https://www.trainingpeaks.com/blog/normalized-power-intensity-factor-training-stress/
Collins, J., Leach, O., Dorff, A., Linde, J., Kofoed, J., Sherman, M.,…Gifford, J. R. (2022). Critical power and work-prime account for variability in endurance training adaptations not captured by V̇o(2max). *J Appl Physiol (1985)*, *133*(4), 986-1000. https://doi.org/10.1152/japplphysiol.00344.2022
Davies, M. J., Benson, A. P., Cannon, D. T., Marwood, S., Kemp, G. J., Rossiter, H. B., & Ferguson, C. (2017). Dissociating external power from intramuscular exercise intensity during intermittent bilateral knee-extension in humans. *J Physiol*, *595*(21), 6673-6686. https://doi.org/10.1113/jp274589
Dawson, B., Fitzsimons, M., Green, S., Goodman, C., Carey, M., & Cole, K. (1998). Changes in performance, muscle metabolites, enzymes and fibre types after short sprint training. *Eur J Appl Physiol Occup Physiol*, *78*(2), 163-169. https://doi.org/10.1007/s004210050402
Edwards, S. (1993). *High performance training and racing , in The Heart Rate Monitor Book*. Feet Fleet Press.
Egan, B., O'Connor, P. L., Zierath, J. R., & O'Gorman, D. J. (2013). Time course analysis reveals gene-specific transcript and protein kinetics of adaptation to short-term aerobic exercise training in human skeletal muscle. *PLoS One*, *8*(9), e74098. https://doi.org/10.1371/journal.pone.0074098
Fiorenza, M., Gunnarsson, T. P., Hostrup, M., Iaia, F. M., Schena, F., Pilegaard, H., & Bangsbo, J. (2018). Metabolic stress-dependent regulation of the mitochondrial biogenic molecular response to high-intensity exercise in human skeletal muscle. *J Physiol*, *596*(14), 2823-2840. https://doi.org/10.1113/jp275972
Foster, C., Casado, A., Esteve-Lanao, J., Haugen, T., & Seiler, S. (2022). Polarized Training Is Optimal for Endurance Athletes. *Med Sci Sports Exerc*, *54*(6), 1028-1031. https://doi.org/10.1249/mss.0000000000002871
Gibala, M. J., Little, J. P., van Essen, M., Wilkin, G. P., Burgomaster, K. A., Safdar, A.,…Tarnopolsky, M. A. (2006). Short-term sprint interval versus traditional endurance training: similar initial adaptations in human skeletal muscle and exercise performance. *J Physiol*, *575*(Pt 3), 901-911. https://doi.org/10.1113/jphysiol.2006.112094
Gilmour, G., & Lydiard, A. (2000). *Running With Lydiard*. Meyer & Meyer Sport.
Halson, S. L. (2014). Monitoring training load to understand fatigue in athletes. *Sports Med*, *44 Suppl 2*(Suppl 2), S139-147. https://doi.org/10.1007/s40279-014-0253-z
Hargreaves, M., & Spriet, L. L. (2020). Skeletal muscle energy metabolism during exercise. *Nat Metab*, *2*(9), 817-828. https://doi.org/10.1038/s42255-020-0251-4

Hawley, J. A. (1993). State of the art training guidelines for endurance performance. *New Zealand Coach*, *2*, 14-19.
Hawley, J. A., Tipton, K. D., & Millard-Stafford, M. L. (2006). Promoting training adaptations through nutritional interventions. *J Sports Sci*, *24*(7), 709-721. https://doi.org/10.1080/02640410500482727
Hazell, T. J., MacPherson, R. E. K., Gravelle, B. M. R., & Lemon, P. W. R. (2010). 10 or 30-s sprint interval training bouts enhance both aerobic and anaerobic performance. *European Journal of Applied Physiology*, *110*(1), 153-160. https://doi.org/10.1007/s00421-010-1474-y
Hellard, P., Avalos, M., Millet, G., Lacoste, L., Barale, F., & Chatard, J. C. (2005). Modeling the residual effects and threshold saturation of training: a case study of Olympic swimmers. *J Strength Cond Res*, *19*(1), 67-75. https://doi.org/10.1519/14853.1
Hickson, R. C., Bomze, H. A., & Holloszy, J. O. (1977). Linear increase in aerobic power induced by a strenuous program of endurance exercise. *J Appl Physiol Respir Environ Exerc Physiol*, *42*(3), 372-376. https://doi.org/10.1152/jappl.1977.42.3.372
Hill, D. W. (1993). The critical power concept. A review. *Sports Med*, *16*(4), 237-254. https://doi.org/10.2165/00007256-199316040-00003
Holloszy, J. O., & Coyle, E. F. (1984). Adaptations of skeletal muscle to endurance exercise and their metabolic consequences. *J Appl Physiol Respir Environ Exerc Physiol*, *56*(4), 831-838. https://doi.org/10.1152/jappl.1984.56.4.831
Hughson, R. L., Orok, C. J., & Staudt, L. E. (1984). A high velocity treadmill running test to assess endurance running potential. *Int J Sports Med*, *5*(1), 23-25. https://doi.org/10.1055/s-2008-1025875
Impellizzeri, F. M., Marcora, S. M., & Coutts, A. J. (2019). Internal and External Training Load: 15 Years On. *Int J Sports Physiol Perform*, *14*(2), 270-273. https://doi.org/10.1123/ijspp.2018-0935
Inglis, E. C., Iannetta, D., Rasica, L., Mackie, M. Z., Keir, D. A., MacInnis, M. J., & Murias, J. M. (2024). Heavy-, Severe-, and Extreme-, but not Moderate-Intensity Exercise Increase V̇o2max and Thresholds after 6 Weeks of Training. *Med Sci Sports Exerc*. https://doi.org/10.1249/mss.0000000000003406
Issurin, V. B. (2010). New horizons for the methodology and physiology of training periodization. *Sports Med*, *40*(3), 189-206. https://doi.org/10.2165/11319770-000000000-00000
Jones, A. M., Burnley, M., Black, M. I., Poole, D. C., & Vanhatalo, A. (2019). The maximal metabolic steady state: redefining the 'gold standard'. *Physiol Rep*, *7*(10), e14098. https://doi.org/10.14814/phy2.14098
Jones, A. M., & Vanhatalo, A. (2017). The 'Critical Power' Concept: Applications to Sports Performance with a Focus on Intermittent High-Intensity Exercise. *Sports Med*, *47*(Suppl 1), 65-78. https://doi.org/10.1007/s40279-017-0688-0
Kiely, J. (2010). New horizons for the methodology and physiology of training periodization: block periodization: new horizon or a false dawn? *Sports Med*, *40*(9), 803-805; author reply 805-807. https://doi.org/10.2165/11535130-000000000-00000
Kontro, H., Mastracci, A., Cheung, S., & MacInnis, M. J. (2024). Maximum Power Available: An Important Concept for Prediction of Task Failure and Improved Estimation of Training Loads in Cycling. *Journal of Science and Cycling*, *13*(2), 7-9.

Larina, I. M., Nosovsky, A. M., & Rusanov, V. B. (2022). Holism and Reductionism in Physiology. *Human Physiology*, *48*(3), 346-354. https://doi.org/10.1134/S036211972201008X
Laursen, P. B. (2010). Training for intense exercise performance: high-intensity or high-volume training? *Scand J Med Sci Sports*, *20 Suppl 2*, 1-10. https://doi.org/10.1111/j.1600-0838.2010.01184.x
Linossier, M. T., Denis, C., Dormois, D., Geyssant, A., & Lacour, J. R. (1993). Ergometric and metabolic adaptation to a 5-s sprint training programme. *Eur J Appl Physiol Occup Physiol*, *67*(5), 408-414. https://doi.org/10.1007/bf00376456
Lucia, A., Hoyos, J., Santalla, A., Earnest, C., & Chicharro, J. L. (2003). Tour de France versus Vuelta a España: which is harder? *Med Sci Sports Exerc*, *35*(5), 872-878. https://doi.org/10.1249/01.mss.0000064999.82036.b4
MacInnis, M., Egan, B., & Gibala, M. (2022). The Effect of Training on Skeletal Muscle and Exercise Metabolism. In (pp. 215-242). https://doi.org/10.1007/978-3-030-94305-9_10
Millet, G. P., Candau, R. B., Barbier, B., Busso, T., Rouillon, J. D., & Chatard, J. C. (2002). Modelling the transfers of training effects on performance in elite triathletes. *Int J Sports Med*, *23*(1), 55-63. https://doi.org/10.1055/s-2002-19276
Monod, H., & Scherrer, J. (1965). The work capacity of a synergic muscular group. *Ergonomics*, *8*(3), 329-338. https://doi.org/10.1080/00140136508930810
Montalvo, S., Gomez, M., Lozano, A., Arias, S., Rodriguez, L., Morales-Acuna, F., & Gurovich, A. N. (2022). Differences in Blood Flow Patterns and Endothelial Shear Stress at the Carotid Artery Using Different Exercise Modalities and Intensities. *Front Physiol*, *13*, 857816. https://doi.org/10.3389/fphys.2022.857816
Moritani, T., Nagata, A., deVries, H. A., & Muro, M. (1981). Critical power as a measure of physical work capacity and anaerobic threshold. *Ergonomics*, *24*(5), 339-350. https://doi.org/10.1080/00140138108924856
Morton, R. H. (1990). Modelling human power and endurance. *J Math Biol*, *28*(1), 49-64. https://doi.org/10.1007/bf00171518
Morton, R. H. (1996). A 3-parameter critical power model. *Ergonomics*, *39*(4), 611-619. https://doi.org/10.1080/00140139608964484
Morton, R. H. (2006). The critical power and related whole-body bioenergetic models. *Eur J Appl Physiol*, *96*(4), 339-354. https://doi.org/10.1007/s00421-005-0088-2
Morton, R. H., Fitz-Clarke, J. R., & Banister, E. W. (1990). Modeling human performance in running. *J Appl Physiol (1985)*, *69*(3), 1171-1177. https://doi.org/10.1152/jappl.1990.69.3.1171
Mujika, I., Busso, T., Lacoste, L., Barale, F., Geyssant, A., & Chatard, J. C. (1996). Modeled responses to training and taper in competitive swimmers. *Med Sci Sports Exerc*, *28*(2), 251-258. https://doi.org/10.1097/00005768-199602000-00015
Mølmen, K. S., Almquist, N. W., & Skattebo, Ø. (2025). Effects of Exercise Training on Mitochondrial and Capillary Growth in Human Skeletal Muscle: A Systematic Review and Meta-Regression. *Sports Med*, *55*(1), 115-144. https://doi.org/10.1007/s40279-024-02120-2
Nummela, A., Hynynen, E., Kaikkonen, P., & Rusko, H. (2010). Endurance performance and nocturnal HRV indices. *Int J Sports Med*, *31*(3), 154-159. https://doi.org/10.1055/s-0029-1243221

Parolin, M. L., Chesley, A., Matsos, M. P., Spriet, L. L., Jones, N. L., & Heigenhauser, G. J. (1999). Regulation of skeletal muscle glycogen phosphorylase and PDH during maximal intermittent exercise. *Am J Physiol*, *277*(5), E890-900. https://doi.org/10.1152/ajpendo.1999.277.5.E890

Poole, D. C., Burnley, M., Vanhatalo, A., Rossiter, H. B., & Jones, A. M. (2016). Critical Power: An Important Fatigue Threshold in Exercise Physiology. *Med Sci Sports Exerc*, *48*(11), 2320-2334.

Poole, D. C., Ward, S. A., Gardner, G. W., & Whipp, B. J. (1988). Metabolic and respiratory profile of the upper limit for prolonged exercise in man. *Ergonomics*, *31*(9), 1265-1279. https://doi.org/10.1080/00140138808966766

Roberts, A. D., Billeter, R., & Howald, H. (1982). Anaerobic muscle enzyme changes after interval training. *Int J Sports Med*, *3*(1), 18-21. https://doi.org/10.1055/s-2008-1026055

Ross, A., & Leveritt, M. (2001). Long-term metabolic and skeletal muscle adaptations to short-sprint training: implications for sprint training and tapering. *Sports Med*, *31*(15), 1063-1082. https://doi.org/10.2165/00007256-200131150-00003

Sanders, D., Abt, G., Hesselink, M. K. C., Myers, T., & Akubat, I. (2017). Methods of Monitoring Training Load and Their Relationships to Changes in Fitness and Performance in Competitive Road Cyclists. *Int J Sports Physiol Perform*, *12*(5), 668-675. https://doi.org/10.1123/ijspp.2016-0454

Scheuer, J., & Tipton, C. M. (1977). Cardiovascular adaptations to physical training. *Annu Rev Physiol*, *39*, 221-251. https://doi.org/10.1146/annurev.ph.39.030177.001253

Skiba, P. F., Fulford, J., Clarke, D. C., Vanhatalo, A., & Jones, A. M. (2015). Intramuscular determinants of the ability to recover work capacity above critical power. *Eur J Appl Physiol*, *115*(4), 703-713. https://doi.org/10.1007/s00421-014-3050-3

Thomas, S. G., Cunningham, D. A., Rechnitzer, P. A., Donner, A. P., & Howard, J. H. (1985). Determinants of the training response in elderly men. *Med Sci Sports Exerc*, *17*(6), 667-672. https://doi.org/10.1249/00005768-198512000-00008

Thomas, T. (2017). *Training with TSS vs. hrTSS: What's the difference?* Retrieved 24/11/2020 from https://www.trainingpeaks.com/blog/training-with-tss-vs-hrtss-whats-the-difference/

Vanhatalo, A., Doust, J. H., & Burnley, M. (2007). Determination of critical power using a 3-min all-out cycling test. *Med Sci Sports Exerc*, *39*(3), 548-555. https://doi.org/10.1249/mss.0b013e31802dd3e6

Vermeire, Van de Casteele, F., Gosseries, M., Bourgois, J. G., Ghijs, M., & Boone, J. (2021). The Influence of Different Training Load Quantification Methods on the Fitness-Fatigue Model. *Int J Sports Physiol Perform*, *16*(9), 1261-1269. https://doi.org/10.1123/ijspp.2020-0662

Vermeire, K., Ghijs, M., Bourgois, J. G., & Boone, J. (2022). The Fitness-Fatigue Model: What's in the Numbers? *Int J Sports Physiol Perform*, *17*(5), 810-813. https://doi.org/10.1123/ijspp.2021-0494

Vinetti, G., Taboni, A., Bruseghini, P., Camelio, S., D'Elia, M., Fagoni, N.,…Ferretti, G. (2019). Experimental validation of the 3-parameter critical power model in cycling. *Eur J Appl Physiol*, *119*(4), 941-949. https://doi.org/10.1007/s00421-019-04083-z

Viru, A., & Viru, M. (2000). *"Nature of training effects". In: Exercise and Sport Science* (Vol. 67-95). Lippincott Williams & Williams.

Wallace, L. K., Slattery, K. M., & Coutts, A. J. (2014). A comparison of methods for quantifying training load: relationships between modelled and actual training responses. *European Journal of Applied Physiology*, *114*(1), 11-20. https://doi.org/10.1007/s00421-013-2745-1

Weatherwax, R. M., Harris, N. K., Kilding, A. E., & Dalleck, L. C. (2019). Incidence of V˙O2max Responders to Personalized versus Standardized Exercise Prescription. *Med Sci Sports Exerc*, *51*(4), 681-691. https://doi.org/10.1249/mss.0000000000001842

Wood, R. E., Hayter, S., Rowbottom, D., & Stewart, I. (2005). Applying a mathematical model to training adaptation in a distance runner. *Eur J Appl Physiol*, *94*(3), 310-316. https://doi.org/10.1007/s00421-005-1319-2